\documentclass[pra,aps,twocolumn,showpacs,amsmath,amssymb,floatfix,10pt]{revtex4-1}
\input{epsf}

\usepackage{graphicx}
\usepackage{dcolumn}
\usepackage{bm}

\begin{document}
\title{Quantum-limited measurement of magnetic-field gradient with entangled atoms}
\author{H. T. Ng}
\affiliation{Center for Quantum Information, Institute for Interdisciplinary Information Sciences, Tsinghua University, Beijing 100084, P. R. China}

\date{\today}

\begin{abstract}
We propose a method to detect the microwave magnetic-field gradient by using a pair of
entangled two-component Bose-Einstein condensates. We consider the two spatially separated 
condensates to be coupled to the two different magnetic fields. The magnetic-field 
gradient can be determined by measuring the variances of population differences and 
relative phases between the two-component condensates in two wells.
The precision of measurement can reach the Heisenberg
limit.  We study the effects of one-body and two-body atom losses on the detection.
We find that the entangled atoms can outperform the uncorrelated atoms in
probing the magnetic fields in the presence of atom losses.  
The effect of atom-atom interactions is also discussed.
\end{abstract}


\maketitle


\section{Introduction}
Probing the magnetic field \cite{Budker} is important in different
areas of science such as physical science \cite{Greenberg} and 
biomedical science \cite{Hamalainen}, etc.
Recently, ultracold atoms have been used for
detecting magnetic field \cite{Wildermuth,Vengalattore,Bohi}
due to long coherence times \cite{Harber,Treutlein} and negligible Doppler
broadening. In addition, coherent collisions between
atoms lead to nonlinear interactions which can be used for
generate quantum entanglement \cite{Horodecki}. In fact, entanglement is a 
useful resource \cite{Giovannetti} for enhancing the accuracy 
of precision measurements. The measurements beyond the standard 
quantum limit have been recently demonstrated \cite{Gross,Riedel} 
by using entangled atomic Bose-Einstein condensates (BECs).

\begin{figure}[ht]
\centering
\includegraphics[height=5.5cm]{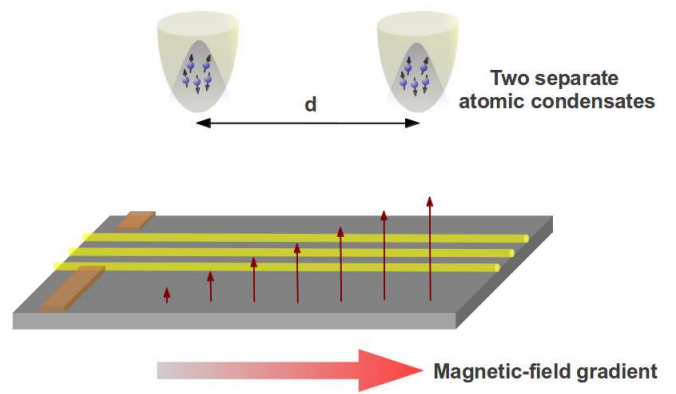}
\caption{ \label{Fig1} (Color online) Schematic
of two atomic condensates being 
placed above a surface which produces a magnetic-field
gradient. The two trapped condensates are separate with a 
distance $d$.  The atoms are coupled to the
magnetic fields through their magnetic dipoles.
 }
\end{figure}

In this paper, we propose a method to detect the microwave magnetic-field
gradient by using two spatially separated condensates of ${}^{87}$Rb atoms
as shown in Fig.~\ref{Fig1}.  
Here we consider the two hyperfine spin states of atoms 
to be coupled to the magnetic fields via their magnetic dipoles \cite{Bohi,Ng1}.
Recently, a BEC has been shown to be transported to a distance about 1 mm
by using a conveyor belt  \cite{Hommelhoff}.  
Therefore, the two separate condensates can be used for measuring
the differences between two magnetic fields at the two different locations.
The magnetic-field gradient can be determined by measuring
the variances of the population differences and the relative phases between 
the two-component condensates in the two different wells.

The sensitivity of the detection can be enhanced by using
entangled atoms \cite{Gross,Riedel,Sorensen}.  Singlet states \cite{Cable}, which are
multi-particle entangled states, have been found useful for 
detecting the magnetic-field gradient \cite{Lanz}. The accuracy of 
measurement can attain the Heisenberg limit  \cite{Cable,Lanz}. In this paper, we discuss
how to produce the singlet state of two spatially 
separated BECs by using entangled tunneling \cite{Ng2}
and appropriately applying the relative phase shifts between
the atoms. It is necessary to manipulate the tunneling
couplings and atom-atom interactions of the condensates
in a double well.  These have been shown in recent
experiments \cite{Albiez,Folling,Esteve}.

However, the performance of detection can be affected by the atom losses
of the condensates \cite{Cooper1,Cooper2,Rey}. In fact, the two-body atom losses \cite{Mertes} are dominant 
in two-component condensates. We study the effects of one-body and two-body losses
on the measurements.  We find that the entangled atoms can give better
performances than using uncorrelated atoms in detecting the magnetic fields if the 
loss rates of atoms are much weaker than the coupling strength of the field gradient. 
Apart from atom losses, the effect of atom-atom interactions 
on the performance of this detection is also important \cite{Grond,Tikhonenkov}.
Here we show that the magnetic-field gradient can be 
estimated if the nonlinear interactions are sufficiently weak.  The accuracy of
the detection will be reduced when the strength of nonlinear interactions
becomes strong. But this can be minimized by either using Feshbach 
resonance \cite{Gross} or state-dependent trap \cite{Riedel}.

\section{System}
We consider two spatially separated Bose-Einstein condensates (BECs) of ${}^{87}$Rb atoms,
where each atom has two hyperfine levels $|e\rangle=|F=2,m_F'\rangle$ and $|g\rangle=|F=1,m_F=-1\rangle$ \cite{Bohi}. 
Here the magnetic number $m_F'$ of the upper hyperfine level can be $-2$, $-1$ or $0$.  
This upper state $|e\rangle$ can be carefully chosen for which the polarization of 
the magnetic field is to be detected \cite{Bohi}.
The two BECs are placed above a surface which generates a 
magnetic-field gradient as shown in Fig.~\ref{Fig1}. 
The two condensates are coupled to the 
two different magnetic fields via their magnetic dipoles \cite{Bohi,Ng1}.

We adopt the two-mode approximation \cite{Milburn} to describe 
the atoms in deep potential wells.  
The Hamiltonian $H_{0}$ can be written as \cite{Ng2} 
\begin{eqnarray}
 H_{0}&=&-\frac{\hbar}{2}({E^e_J}e^\dag_Le_R+{E^g_J}g^\dag_Lg_R+{\rm H.c})\nonumber\\
 &&+{\hbar}\sum_{\alpha={L,R}}(U_{ee}{n}^2_{e_\alpha}+2U_{eg}{n}_{e_\alpha}n_{g_\alpha}+U_{gg}n^2_{g_\alpha}),
\end{eqnarray}
where $e_\alpha(g_\alpha)$ and $n_{e_\alpha}(n_{g_\alpha})$ are the annihilation and 
number operators of the atoms in the 
state $|e\rangle(|g\rangle)$ in the left and right potential wells, respectively. 
The parameters $E^e_J(E^g_J)$ and $U_{ee}(U_{gg})$ 
are the tunneling strength between the two wells and the atom-atom interaction
strength, for the states $|e\rangle(|g\rangle)$, and $U_{eg}$ is
the interaction strength between the atoms in the two different components.

We consider the atoms to be resonantly coupled to the microwave
magnetic fields. The transition frequencies of hyperfine 
states can be tuned by a static magnetic field \cite{Bohi}.
Here the other hyperfine transitions can be ignored due to large detuning \cite{Bohi}.  
The Hamiltonian $H_I$, describes the internal states and their interactions
between the magnetic fields, is given by \cite{Bohi}
\begin{equation}
\label{Hamint}
 H_{I}=\hbar\sum_{\alpha=L,R}\Big[{\omega_e}n_{e\alpha}+{\omega_g}n_{g\alpha}
+\frac{\Omega_\alpha}{2}(e^{i\omega{t}}e^\dag_{\alpha}g_{\alpha}+{\rm H.c.})\Big],
\end{equation}
where $\omega_e$ and $\omega_g$ are the frequencies of the atoms in the states
$|e\rangle$ and $|g\rangle$, respectively, and $\omega$ is the frequency of
the magnetic field. The parameter $\Omega_L(\Omega_R$) is the coupling 
strength between the atoms and magnetic field $B_{L}(B_R$) in the left(right)
potential well.

We work in the interaction picture by performing the unitary transformation as
\begin{eqnarray}
U(t)&=&\exp\Bigg[-it\sum_{\alpha=L,R}({\omega}n_{e\alpha}+\omega_gn_{g\alpha})\Bigg].
\end{eqnarray}
The transformed Hamiltonian becomes
\begin{eqnarray}
\label{Hamint}
 H_{I}&=&\hbar\sum_{\alpha=L,R}\Big[{\Delta}n_{e\alpha}+
\frac{\Omega_\alpha}{2}(e^\dag_{\alpha}g_{\alpha}+{\rm H.c.})\Big],
\end{eqnarray}
where $\Delta=\omega_e-\omega$ is the detuning between the atoms
and the magnetic field.

If the two wells are separate with a large distance, then the 
tunneling strengths are effectively turned off, i.e., $E^e_J=E^g_J=0$.
Here we consider the number of atoms in each trap to be equal to $N/2$, where $N$
is the total number of atoms. For convenience,  the system can be expressed 
in terms of angular momentum operators as \cite{Sakurai}:
\begin{eqnarray}
J_{\alpha{x}}&=&\frac{1}{2}(e^\dag_\alpha{g_\alpha}+g^\dag_\alpha{e_\alpha}),\\
J_{\alpha{y}}&=&\frac{1}{2i}(e^\dag_\alpha{g_\alpha}-g^\dag_\alpha{e_\alpha}),\\
J_{\alpha{z}}&=&\frac{1}{2}(e^\dag_\alpha{e_\alpha}-g^\dag_\alpha{g_\alpha}),
\end{eqnarray}
where $\alpha=L,R$.  
The Hamiltonians $H_0$ and $H_I$ are rewritten as
\begin{eqnarray}
\label{Ham0}
H_0&=&\hbar\sum_{\alpha={L,R}}\Big[\frac{1}{2}(U_{ee}-U_{gg}){N}J_{\alpha{z}}+{\chi}J^2_{\alpha{z}}\Big],\\
H_I&=&{\hbar}\sum_{\alpha=L,R}(\Delta{J}_{\alpha{z}}+\Omega_\alpha{J}_{\alpha{x}}),
\end{eqnarray}
where $\chi=U_{ee}+U_{gg}-2U_{eg}$.
We have omitted a constant term $\hbar(U_{ee}N^2+U_{gg}N^2+2U_{eg}N^2+4{\Delta}N)/8$.

\section{Detection of magnetic-field gradient}
We present a scheme for detecting the magnetic-field gradient by using a pair of
entangled BECs. First, it is necessary to generate
the entanglement between two spatially separated condensates.  Then, one of
the condensates can be brought to another place for detection and the atoms
are coupled to the magnetic fields.  The magnetic-field gradient can be
estimated by measuring the variances of the population differences and 
relative phases between the condensates in the two internal states.  
The procedure of this detection scheme is
described in the following:
 
\subsection{Generation of entangled states}
To enhance the sensitivity of detection, it is necessary to
generate the entanglement between the two separate condensates. 
We consider the condensates 
to be prepared in an entangled state which is given by \cite{Cable}
\begin{eqnarray}
\label{entinput}
|\Psi_{\rm in}\rangle&=&\frac{1}{\sqrt{2j+1}}\sum_{m}(-1)^m|j,m\rangle_L|j,-m\rangle_R,
\end{eqnarray}
where $|j,m\rangle_\alpha$ is an eigenstate of angular momentum 
operator $J_{\alpha{z}}$ for $\alpha=L,R$, and $j=N/4$.  
The spin states of the two condensates are anti-correlated \cite{Ng2}.

\begin{figure}[ht]
\centering
\includegraphics[height=8cm]{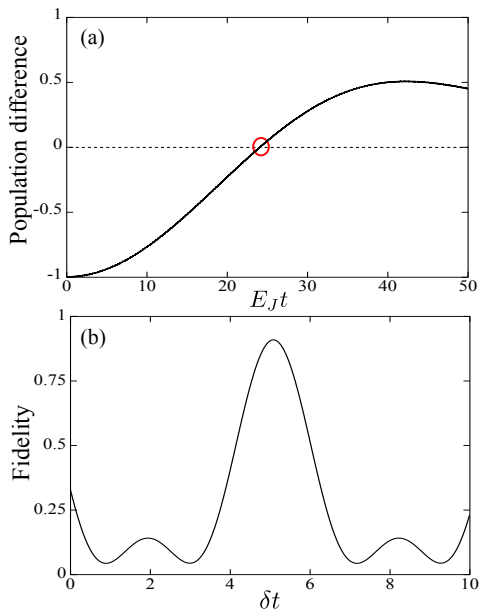}
\caption{ \label{fig2} (Color online)   (a) Time evolution
of the population difference between two wells versus time, 
for the atoms in the state $|g\rangle$. (b) Fidelity between 
the input state $|\Psi_{\rm in}\rangle$ and 
the state $|\psi(t)\rangle$ versus time.  The red
circle denotes the zero population difference between the two 
wells at the time $t^*$.
The parameters are used: $N=4$, $U=10E_J$, $\delta_L=\delta$
and $\delta_R=0$. 
 }
\end{figure}

Now we discuss how to generate the entangled state $|\Psi_{\rm in}\rangle$
in Eq.~(\ref{entinput}). Initially, the atoms in the two different internal states
$|e\rangle$ and $|g\rangle$ are confined in the different potential wells, respectively,
where each condensate has an equal number of atoms, $N/2$. 
The intra- and inter-component interaction strengths are
tuned to be the same, i.e.,  $U\!\approx\!U_{ee}\!\approx\!{U_{gg}}\!\approx\!U_{eg}/2$.
The tunneling strengths $E_J\!\approx\!{E^e_J}\!\approx\!{E^g_J}$ are much weaker than 
the atomic interaction strengths $U$.
The entanglement between the atoms in the two components  
can be dynamically produced via the process of tunneling \cite{Ng2}.  
In this process, the atoms in the two different internal states can
tunnel in pair due to the strong atom-atom interactions.
The entangled state can then be generated as \cite{Ng2}
\begin{equation}
\label{qct1}
|\Psi(t^*)\rangle\approx\sum_n{c_n}|n\rangle_{g_L}|N/2-n\rangle_{g_R}|N/2-n\rangle_{e_L}|n\rangle_{e_R},
\end{equation}
where $c_n$ is the probability amplitude.  When the population 
difference between the wells for each component condensate
is about zero at the time $t^*$, the probability
amplitude $c_n$ is approximately equal to $1/\sqrt{N/2+1}$. 
Then, the tunneling can be effectively switched off by adiabatically separating the 
two wells.  Since the number of atoms in each well is equal to each other, the two-component condensate
in each trap can be effectively described by an angular momentum system.
Therefore, the state $|\Psi(t^*)\rangle$ can be rewritten as \cite{Ng2}
\begin{figure}[ht]
\centering
\includegraphics[height=5cm]{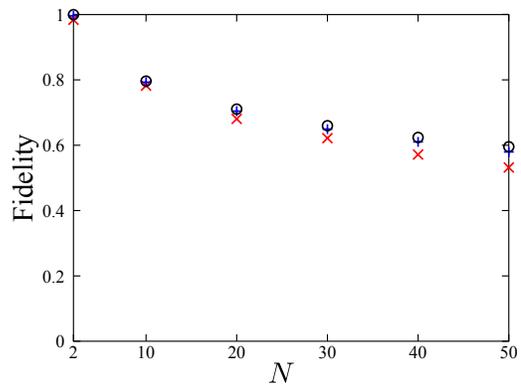}
\caption{ \label{fig3} (Color online)   Fidelity between the input state $|\Psi_{\rm in}\rangle$ 
and the entangled state $|\tilde{\Psi}\rangle$ versus the total number $N$ of atoms.
The fidelities are shown with the different parameters for generating $|\tilde{\Psi}\rangle$: 
$U=5E_J$ (red cross), $10E_J$ (blue plus) and $50E_J$ (black empty circle), respectively.
}
\end{figure}
\begin{eqnarray}
\label{qct2}
|\Psi(t^*)\rangle\!&\approx&\!\frac{1}{\sqrt{2j+1}}\!\sum_m\!|n\rangle_{g_L}|N/2-n\rangle_{e_L}|N/2-n\rangle_{g_R}|n\rangle_{e_R},\nonumber\\
\label{entunnel1}
&\approx&\frac{1}{\sqrt{2j+1}}\sum_m|j,m\rangle_L|j,-m\rangle_R,
\end{eqnarray}
where $|j,m\rangle_L=|n\rangle_{g_L}|N/2-n\rangle_{e_L}$ and $|j,-m\rangle_R=|N/2-n\rangle_{g_R}|n\rangle_{e_R}$
are the eigenstates of $J_{Lz}$ and $J_{Rz}$, respectively.

In Fig.~\ref{fig2}(a), we plot the population difference between the two wells versus time, where
the atoms are in $|g\rangle$. A red empty circle 
denotes the time $t^*$ for which the population difference is equal to zero.
The atoms in $|g\rangle$ can tunnel to the other well even if the interaction
strengths $U$ are much stronger than $E_J$.

The entangled state $|\Psi(t^*)\rangle$ in Eq.~(\ref{qct2})
differs from the entangled state $|\Psi_{\rm in}\rangle$ in Eq.~(\ref{entinput}) 
in the relative phases between the atoms.
A relative phase shift can be accumulated by turning on the interaction which
can be described by the Hamiltonian $H_{\rm rp}$ as
\begin{eqnarray}
H_{\rm rp}&=&\hbar(\delta_{L}J_{L{z}}+\delta_{R}J_{R{z}}),
\end{eqnarray}
where $\delta_{L}$ is not equal to $\delta_{R}$. 
The state $|\psi(t)\rangle$ can be produced as
\begin{eqnarray}
\label{relphase}
|\psi(t)\rangle&=&\exp(-iH_{\rm rp}t)|\Psi(t^*)\rangle.
\end{eqnarray}
This interaction can be made by
controlling the strength $U_{ee}$ and $U_{gg}$ in Eq.~(\ref{Ham0}) in one of 
the potential wells.
Note that this interaction will not change the population difference
between the two-component condensates. Therefore,
the quantum numbers $m$ in Eq.~(\ref{entunnel1}) 
remain unchanged during the interaction.
After turning on the interaction for a specific time,
the required entangled state $|\tilde{\Psi}\rangle$ 
can then be produced, where $|\tilde{\Psi}\rangle$
is the state which has the maximum fidelity \cite{Uhlmann} between 
$|\Psi_{\rm in}\rangle$ and $|\psi(t)\rangle$.

We then study the fidelity between the states $|\Psi_{\rm in}\rangle$
and $|\psi(t)\rangle$.
In Fig.~\ref{fig2}(b), we plot the fidelity, $|\langle\Psi_{\rm in}|\psi(t)\rangle|^2$, 
is plotted versus the 
time, where $\delta_L=\delta$ and $\delta_R=0$.  
The fidelity varies with the time $t$ as shown in Fig.~\ref{fig2}(b). 
The highest fidelity can exceed 0.9.

In addition, we examine the fidelity between the states $|\Psi_{\rm in}\rangle$ 
and $|\tilde{\Psi}\rangle$ for the different numbers $N$ of atoms in Fig.~\ref{fig3}.
As $N$ increases, the fidelities decreases.
However, the higher fidelity can be obtained with a higher ratio of $U$
to $E_J$.

\subsection{Coupling to the magnetic field}
We consider the atoms to be coupled to the magnetic field at
resonance, i.e., $\Delta=0$ and set $U_{ee}=U_{gg}$. 
Here we assume that the strengths of atom-atom interactions are
much weaker than the coupling strengths $\Omega_L$
and $\Omega_R$, and therefore they are ignored here. The effect of
the atom-atom interactions will be discussed later.
The Hamiltonian $H_I$ reads
\begin{eqnarray}
H_I&=&{\hbar}(\Omega_L{J}_{L{x}}+\Omega_R{J}_{R{x}}).
\end{eqnarray}
The magnetic coupling strengths $\Omega_L$ and $\Omega_R$
are different to each other.  Let us write $\Omega_L=\Omega+\Omega_D/2$
and $\Omega_R=\Omega-\Omega_D/2$. The Hamiltonian $H_I$
can be written as
\begin{eqnarray}
H_1&=&{\hbar}\Omega({J}_{L{x}}+{J}_{R{x}})+\frac{\hbar\Omega_D}{2}({J}_{L{x}}-{J}_{R{x}}).
\end{eqnarray}
This small parameter $\Omega_D$ is to be determined.

\subsection{Read-out process}
The magnetic-field gradient can be estimated by
measuring the variance $\langle\tilde{J}^2_{yz}\rangle=\langle{J^2_{y-}-J^2_{z+}}\rangle$,
where $J_{y-}=J_{Ly}-J_{Ry}$ and $J_{z+}=J_{Lz}+J_{Rz}$. 
Physically speaking, $\langle{J_{\alpha{y}}}\rangle$ and 
$\langle{J_{\alpha{z}}}\rangle$ are the expectation values of the relative phase and population 
difference between the two-component condensates in the potential
well $\alpha$, for $\alpha=L,R$.  
The variance $\langle\tilde{J}^2_{yz}\rangle$ is given by
\begin{eqnarray}
\label{estimator}
\langle{\tilde{J}^2_{yz}}(\phi_D)\rangle&=&\frac{N(N+4)}{12}\cos{(\phi_D)}.
\end{eqnarray}
where $\phi_D=\Omega_Dt$.  Here $\langle{J_{y-}}\rangle$
and $\langle{J_{z+}}\rangle$ are equal to zero
for the input state $|\Psi_{\rm in}\rangle$ in Eq.~(\ref{entinput}).
The expectation value $\langle{\tilde{J}^2_{yz}(\phi_D)}\rangle$ 
is a function of the parameter $\phi_D$.
Therefore, this quantity can be used for
determining the magnetic-field gradient $\Omega_D$.
\begin{figure}[ht]
\centering
\includegraphics[height=5.5cm]{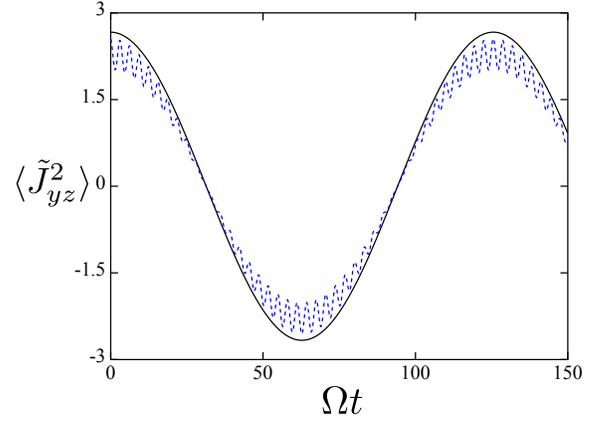}
\caption{ \label{fig4} (Color online) The variances 
$\langle\tilde{J}^2_{yz}\rangle$ are plotted versus time, for $N=4$ and 
$\Omega_D=0.05\Omega$.  
The black solid and blue dotted lines are shown by using the two different 
initial states $|\Psi_{\rm in}\rangle$
and $|\tilde{\Psi}\rangle$, respectively, where $U=10E_J$ is used for producing
$|\tilde{\Psi}\rangle$. 
}
\end{figure}
\begin{figure}[ht]
\centering
\includegraphics[height=5.5cm]{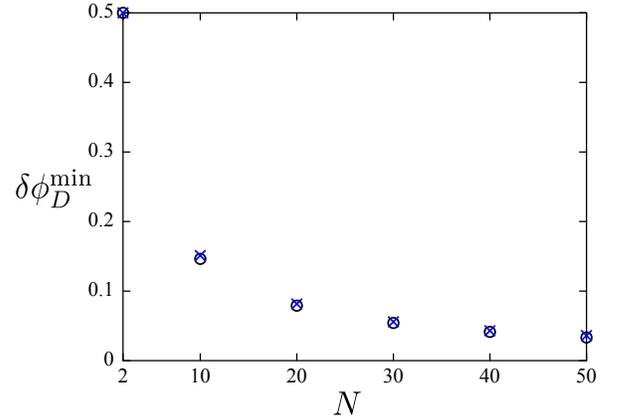}
\caption{ \label{fig5} (Color online)   The minimum of
the uncertainties $\delta\phi^{\rm min}_D$ versus 
the total number $N$ of atoms, for $\Omega_D=0.05\Omega$.
The empty circles and blue crosses denote the system with
the two different initial states $|\Psi_{\rm in}\rangle$
and $|\tilde{\Psi}\rangle$, respectively, where $U=10E_J$ 
is used for producing the state $|\tilde{\Psi}\rangle$.
 }
\end{figure}
In Fig.~\ref{fig4}, we plot the variance $\langle\tilde{J}^2_{yz}\rangle$
versus the time, for the two different initial states $|\Psi_{\rm in}\rangle$
and $|\tilde{\Psi}\rangle$, respectively.  The variances 
$\langle\tilde{J}^2_{yz}\rangle$ oscillate with the frequency 
$\phi_D$, for these two initial states.  But the variance $\langle\tilde{J}^2_{yz}\rangle$
shows some small-amplitude fluctuations in the slow oscillations if the initial state 
$|\tilde{\Psi}\rangle$ is used.

\subsection{Sensitivity of detection}
The magnetic-field gradient can be estimated from the variance $\langle\tilde{J}^2_{yz}\rangle$.
The uncertainty of the parameter $\phi_D$ is given by
\begin{eqnarray}
\delta{\phi}_D&=\dfrac{\Delta{\tilde{J}^2_{yz}}}{|\partial\langle{\tilde{J}^2_{yz}}\rangle/\partial\phi_D|},
\end{eqnarray}
where $\Delta{\tilde{J}^2_{yz}}=\sqrt{\langle\tilde{J}^4_{yz}\rangle-\langle{\tilde{J}^2_{yz}}\rangle^2}$.
The uncertainty $\delta\phi_D$ can be found as 
\begin{equation}
\delta\phi_D=\sqrt{\frac{15\sin^2(\phi_D)+(N-2)(N+6)\cos^2({\phi_D)}}{5N(N+4)\sin^2(\phi_D)}}.
\end{equation}
At the time $t=\pi/2\Omega_D$, the minimum uncertainty $\delta\phi^{\rm min}_D$ is 
\begin{eqnarray}
\delta\phi^{\rm min}_D&=&\sqrt{\frac{3}{N(N+4)}}.
\end{eqnarray}
The uncertainty scales with $1/N$, for large $N$.  Thus, the 
accuracy of measurement can reach
the Heisenberg limit \cite{Giovannetti}.

In Fig.~\ref{fig5}, we plot the minimum uncertainties
$\delta\phi^{\rm min}_D$ versus the total number of 
atom, for the two different initial states $|\Psi_{\rm in}\rangle$
and $|\tilde{\Psi}\rangle$, respectively.  The measurement
using the two different initial states can give the similar 
values of $\delta\phi^{\rm min}_D$.  Therefore, the 
entangled state $|\tilde{\Psi}\rangle$ can provide 
a similar accuracy of the case using the input
state $|\Psi_{\rm in}\rangle$ in Eq.~(\ref{entinput}).

\section{Effect of atom losses}
Now we study the sensitivity of the detection in the presence of
one-body and two-body atom losses.
\begin{figure}[ht]
\centering
\includegraphics[height=9.3cm]{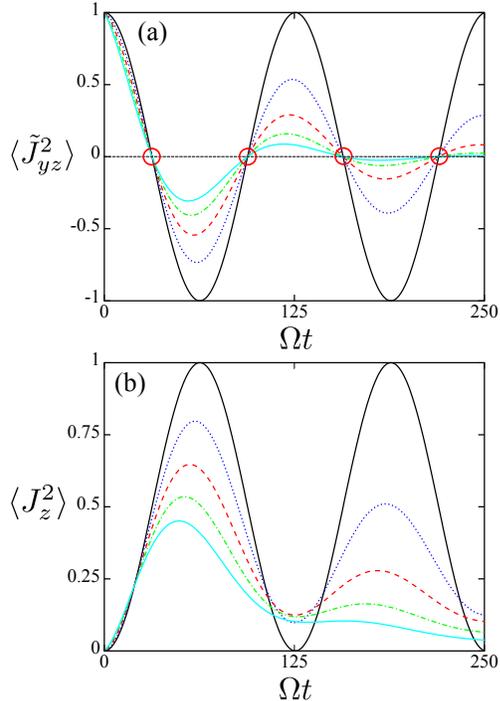}
\caption{ \label{oalossest1} (Color online)   (a)  Variances $\langle\tilde{J}^2_{yz}\rangle$ versus time.
(b) Variances $\langle{J}^2_{z}\rangle$ versus time.  The initial state
is the entangled state $|\Psi_{\rm in}\rangle$ in Eq.~(\ref{entinput}). The parameters are
used: $\Omega_D=0.05\Omega$ and $N=2$.
The different damping rates are shown: $\gamma^o=0$ (black solid line), 
$0.0025\Omega$ (blue dotted line), $0.005\Omega$ (red dashed line), 
$0.0075\Omega$ (green dot-dashed line) and $0.01\Omega$ (cyan solid line),
respectively.  The red empty circles denote the intersection points 
at $\langle\tilde{J}^2_{yz}\rangle=0$. 
 }
\end{figure}
\subsection{One-body atom loss}
Here we study the one-body atom losses by using the phenomenological master
equation \cite{Cooper1,Cooper2}. The master equation, describes one-body atom losses, can be written as \cite{Cooper1,Cooper2}
\begin{equation}
\dot{\rho}=i[\rho,H]+\sum_{\alpha,\beta}\frac{\gamma^{o}_\beta}{2}(2\beta_\alpha\rho\beta^\dag_\alpha-\beta^\dag_\alpha\beta_\alpha\rho-\rho\beta^\dag_\alpha\beta_\alpha),
\end{equation}
where $\gamma^{o}_\beta$ is the damping rate of one-body atom loss, and $\alpha=L,R$ and $\beta=e,g$.

We compare the two estimators $\langle\tilde{J}^2_{yz}\rangle$ and $\langle{J}^2_z\rangle$
for determining the parameter $\phi_D$
in the presence of one-body atom loss, where $J_z=J_{Lz}+J_{Rz}$ is the sum of the
population difference between the two hyperfine spin states of condensates in the two wells.
In Fig.~\ref{oalossest1}(a), we plot the variance $\langle\tilde{J}^2_{yz}\rangle$ versus
time for the different damping rates $\gamma^o=\gamma^o_e=\gamma^o_g$.  

\begin{figure}[ht]
\centering
\includegraphics[height=5.1cm]{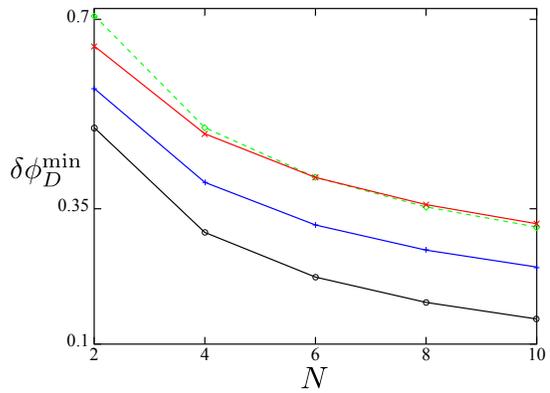}
\caption{ \label{onefig1} (Color online)  The minimum uncertainties $\delta\phi^{\rm min}_D$ 
are plotted versus $N$ for $\Omega_D=0.05\Omega$. 
The different damping rates are shown: $\gamma^o=0$ (black empty circle), 
$0.005\Omega$ (blue plus) and $0.01\Omega$ (red cross), respectively.  
The green diamonds (dashed line) denote the case of using 
uncorrelated atoms with the uncertainty, $1/\sqrt{N}$, for $\gamma^o=0$.
 }
\end{figure}
The initial state is $|\Psi_{\rm in}\rangle$ in Eq.~(\ref{entinput}).
We can see that the variances
$\langle\tilde{J}^2_{yz}\rangle$ intersect at the same point at the time $t=n\pi/2\Omega_D$,
where $n$ is an odd number.  The parameter $\Omega_D$ can be estimated
in the vicinity of these intersection points. In fact, the minimum uncertainty of
the parameter $\Omega_D$ can be obtained at the first intersection point, i.e., $t=\pi/2\Omega_D$.

In Fig.~\ref{oalossest1}(b), the variances $\langle{J}^2_z\rangle$
are plotted versus time, for the different damping rates $\gamma^o$.  
For $\gamma^o=0$, the variance $\langle{J^2_z}\rangle$ can be used
for determining the parameter $\Omega_D$ \cite{Lanz}.
However, the estimators $\langle{J}^2_z\rangle$ do not
intersect at the same point for the different damping rates $\gamma^o$.
Besides, the atom losses cause a shift of the oscillations.  This means
that $\langle{J}^2_z\rangle$ is not a faithful estimator for determining
the parameter $\Omega_D$ in the presence of atom losses.

In Fig.~\ref{onefig1}, we plot the minimum uncertainties $\delta\phi^{\rm min}_D$ 
versus the total number $N$ of atoms, where 
$N$ is up to 10.  
Here the minimum uncertainty 
$\delta\phi^{\rm min}_D$ are obtained at the time $t=\pi/2\Omega_D$.
For comparsion, the case using uncorrelated atoms without any atom loss 
is shown (green diamonds in Fig.~\ref{onefig1}), where the uncertainty is
equal to $1/\sqrt{N}$ \cite{Giovannetti}. 
In Fig.~\ref{onefig1}, the entangled atoms can give a better performance 
than the uncorrelated atoms in detection if the damping rate $\gamma^{o}$ 
is much smaller than $\Omega_D=0.05\Omega$.  
When $\gamma^o=0.01\Omega$ becomes comparable
to $\Omega_D$, the accuracy of the detection is similar to the case using uncorrelated atoms 
as shown in Fig.~\ref{onefig1}.

\subsection{Two-body atom loss}
The phenomenological master equation, describes two-body atom losses, can be written as \cite{Rey}
\begin{eqnarray}
\dot{\rho}&=&\!i[\rho,H]\!+\frac{\gamma^t_{ee}}{2}\!\!\sum_{\alpha=L,R}\!(2e^2_\alpha{\rho}{e^{\dag{2}}_\alpha}-e^{\dag{2}}_\alpha{e}^2_\alpha\rho
-\rho{e}^{\dag{2}}_\alpha{e}^2_\alpha)\nonumber\\
&&+\frac{\gamma^t_{eg}}{2}\!\!\sum_{\alpha=L,R}\!(2e_\alpha{g_\alpha}{\rho}e^\dag_\alpha{g^\dag_\alpha}-e^\dag_\alpha{e}_\alpha{g^\dag_\alpha{g}_\alpha}\rho
-\rho{e}^\dag_\alpha{e}_\alpha{g^\dag_\alpha{g}_\alpha}),\nonumber\\
\end{eqnarray}
where the parameters $\gamma^{t}_{ee}$ and $\gamma^{t}_{eg}$ are the damping rates of 
two-body atom losses for the condensates in the upper internal state $|e\rangle$ and the atoms 
in the two different components, and $\alpha=L,R$.
\begin{figure}[ht]
\centering
\includegraphics[height=9.5cm]{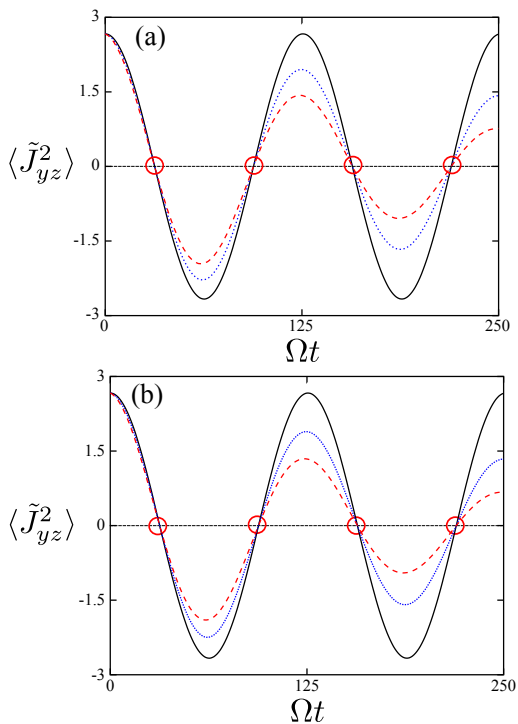}
\caption{ \label{talossest1} (Color online)   Variances $\langle\tilde{J}^2_{yz}\rangle$ versus time.
(a)  The different damping rates are shown: $\gamma^t_{eg}=0$ (black solid line), 
$0.005\Omega$ (blue dotted line) and $0.01\Omega$ (red dashed line),
respectively, and $\gamma^t_{ee}=0$.  
(b)  
The different damping rates are shown: $\gamma^t_{eg}=0$ (black solid line), 
$0.001\Omega$ (blue dotted line) and $0.002\Omega$ (red dashed line),
respectively, and $\gamma^t_{ee}=1.5\gamma^t_{eg}$.  
The initial state
is the input state $|\Psi_{\rm in}\rangle$. The parameters are
used: $\Omega_D=0.05\Omega$ and $N=4$.
The red empty circles denote the intersection points 
at $\langle\tilde{J}^2_{yz}\rangle=0$. 
 }
\end{figure}

In Fig.~\ref{talossest1}(a) and (b), we plot the estimators 
$\langle\tilde{J}^2_{yz}\rangle$ versus time, for the different
damping rates $\gamma^t_{eg}$ of two-body atom losses,
and $\gamma^t_{ee}=0$ in (a) and $\gamma^t_{ee}=1.5\gamma^t_{eg}$ \cite{Mertes} 
in (b), respectively. 
Both of the results show that $\langle\tilde{J}^2_{yz}\rangle$
intersect at the times $n\pi/2\Omega_D$, where $n$ is an odd number.
Therefore, the parameter $\Omega_D$ can be estimated at
the times $n\pi/2\Omega_D$ in the presence of two-body atom losses.

\begin{figure}[ht]
\centering
\includegraphics[height=5.5cm]{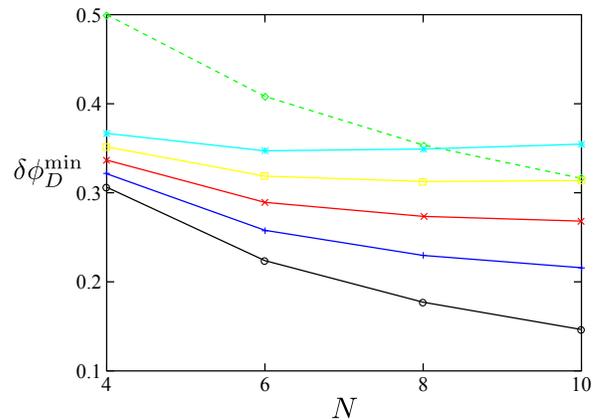}
\caption{ \label{twofig1} (Color online)  The minimum uncertainties $\delta\phi^{\rm min}_D$ 
versus $N$ for $\Omega_D=0.05\Omega$. 
The different damping rates $\gamma^{t}_{eg}$ are shown: 
$\gamma^{t}_{eg}=0$ (black empty circle), $0.0025\Omega$ (blue plus), $0.005\Omega$ (red cross), 
$0.0075\Omega$ (yellow square), $0.01\Omega$ (cyan star) respectively.  
The green diamonds (dashed line) denote the uncertainty, $1/\sqrt{N}$, by using
uncorrelated atoms without any atom loss.
 }
\end{figure}
In Fig.~\ref{twofig1}, we plot the minimum uncertainties $\delta\phi^{\rm min}_D$ versus $N$,
where the minimum uncertainties are taken at the time $t=\pi/2\Omega_D$. The uncertainties from the measurement with 
uncorrelated atoms are shown with green diamonds, where $\gamma^t_{eg}=\gamma^t_{ee}=0$.
The parameters $\delta\phi^{\rm min}_D$ have the different scalings with $N$, 
for the different rates $\gamma^{t}_{eg}$ and $\gamma^{t}_{ee}=0$. 
For small $\gamma^t_{eg}$, 
the entangled atoms can outperform the uncorrelated atoms for detection.
When $\gamma^t_{eg}\geq{0.0075\Omega}$ and $N\geq{8}$, 
the uncertainty $\delta\phi^{\rm min}_D$ does not decrease with $N$. 
To obtain the good performance of the measurements, the damping 
rates $\gamma^t_{eg}$ have to be much smaller than the coupling strength
of the magnetic-field gradient.
\begin{figure}[ht]
\centering
\includegraphics[height=5.5cm]{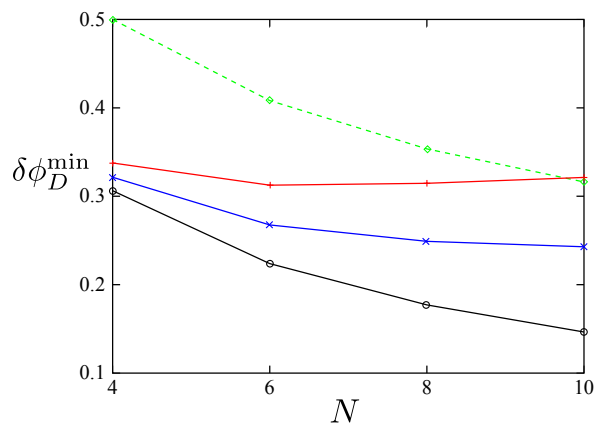}
\caption{ \label{twofig2} (Color online)  The minimum uncertainties $\delta\phi^{\rm min}_D$ 
versus $N$, for $\Omega_D=0.05\Omega$. 
The different damping rates are shown: 
$\gamma^{t}_{eg}=0$ (black empty circle), $\gamma^{t}_{eg}=0.001\Omega$ (blue cross) and 
$\gamma^{t}_{eg}=0.002\Omega$ (red plus), respectively, for
$\gamma^{t}_{ee}=1.5\gamma^{t}_{eg}$.  The green diamonds (dashed line) denote the case of using
uncorrelated atoms with the uncertainty, $1/\sqrt{N}$.
 }
\end{figure}

Then, we study the sensitivity of the detection by including the two-body atom
loss for the atoms in the excited states in $|e\rangle$ \cite{Mertes}.
In Fig.~\ref{twofig2}, we plot the minimum uncertainty $\delta\phi^{\rm min}_D$ versus $N$, where
the minimum uncertainties are taken at the time 
$t=\pi/2\Omega_D$.  Here we set $\gamma^{t}_{ee}=1.5\gamma^{t}_{eg}$ \cite{Mertes}. 
The minimum uncertainty exceeds the case of using uncorrelated atoms when 
$\gamma^{t}_{eg}$ is equal to $0.002\Omega$.
In this case, the two-body atom losses become more detrimental to 
the performance of the detection.

\section{Effect of atom-atom interactions}
\begin{figure}[ht]
\centering
\includegraphics[height=5.5cm]{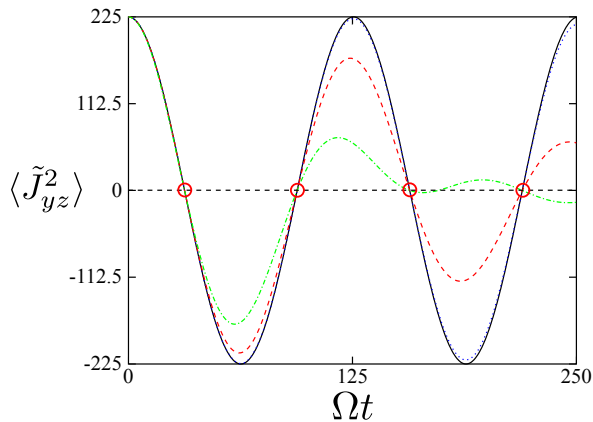}
\caption{ \label{nlintfig1} (Color online)   Time evolution
of variances $\langle\tilde{J}^2_{yz}\rangle$, for
different strengths of atom-atom interactions: $\chi=0$ (black-solid line),
$10^{-4}\Omega$ (blue-dotted line), $5\times{10^{-4}}\Omega$
(red-dashed line) and $0.001\Omega$ (green-dot-dashed line), 
respectively.
The parameters are used: $N=50$ and $\Omega_D=0.05\Omega$.
The red empty circles denote the intersection points at
$\langle\tilde{J}^2_{yz}\rangle=0$.
 }
\end{figure}
We investigate the effect of the atom-atom interactions on 
the detection of magnetic-field gradient. 
In Fig.~\ref{nlintfig1}, we plot the variances
$\langle\tilde{J}^2_{yz}\rangle$ versus time, 
for the different nonlinear interaction strengths $\chi$.  
When $\chi{N}\ll{\Omega}_D$,  $\langle\tilde{J}^2_{yz}(t)\rangle$
are close to each other, for the different strengths $\chi$.
If the nonlinear interaction strength $\chi$ increases,
then the amplitude of oscillations decreases as shown in Fig.~\ref{nlintfig1}. 
In addition, the variances $\langle\tilde{J}^2_{yz}(t)\rangle$, for $\chi{N}{\ll}\Omega_D$, 
almost meet at the same points $\langle\tilde{J}^2_{yz}\rangle=0$ at the times $t=n\pi/2\Omega_D$, 
where $n$ is an odd number.  
At the times $n\pi/2\Omega_D$, these give the minimum 
uncertainty of the parameter $\phi_D$.

\begin{figure}[ht]
\centering
\includegraphics[height=5.5cm]{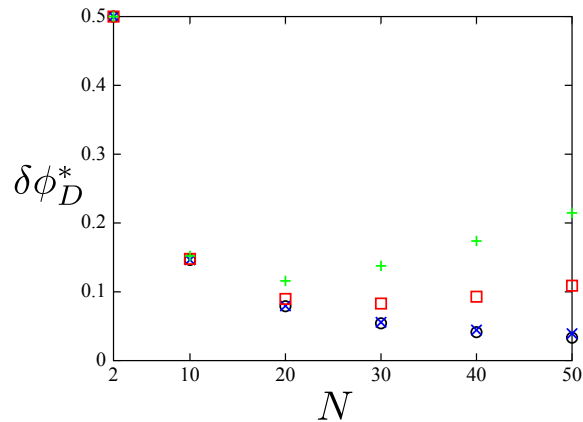}
\caption{ \label{nlintphi1} (Color online)   The uncertainties
$\delta\phi^*_D$ are plotted versus $N$, for $\delta\Omega_D=0.05\Omega$.
The different strengths of atom-atom interactions are shown: 
$\chi=0$ (black empty circle),
$10^{-4}\Omega$ (blue cross), $5\times{10^{-4}}\Omega$
(red square) and $0.001\Omega$ (green plus), 
respectively.
 }
\end{figure}

Then, we investigate the uncertainty $\delta\phi^*_D$ at the time $t=\pi/2\Omega_D$.
In Fig.~\ref{nlintphi1}, we plot the uncertainties $\delta\phi^*_D$ versus
$N$, for the different nonlinear interaction strengths $\chi$.  
The uncertainty $\delta\phi^*_D$ is close to the minimum uncertainty
$\delta\phi^{\rm min}_D$ for $\chi{N}\ll{\Omega_D}$.  When $\chi$ increases, the 
uncertainty $\delta\phi^*_D$ does not decrease for larger $N$ as shown
in Fig.~\ref{nlintphi1}. Therefore, the strong nonlinear interactions limit the 
performance of detection. 

In fact, the effects of nonlinear interactions can be minimized 
by setting $\chi{N}{\ll}\Omega_D$.
The nonlinear interaction strength can be appropriately adjusted by
using Feshbach resonance \cite{Gross} and state-dependent 
trap \cite{Riedel}. 

\section{Discussion}
Let us make some remarks on our method for 
detecting the magnetic-field gradient by using
${}^{87}$Rb atoms. The transition frequency of ${}^{87}$Rb atoms
can be tuned by using an external
static magnetic field \cite{Bohi}, the range of the frequencies of the detected
magnetic field is about a few GHz to 10 GHz \cite{Bohi,Steck}.  

Next, we roughly estimate the
magnitude of the magnetic-field gradient which can be probed by using the condensates.  
Indeed, the measurement is mainly limited by the atom loss rate of the condensates.
The main source comes from two-body atom losses \cite{Mertes}.  The two-body loss
rate $(\gamma^t_{eg}+\gamma^t_{ee})N/V$, where 
$\gamma^t_{eg}(\gamma^t_{ee})\sim{10^{-13}}{\rm cm}^3{\rm s}^{-1}$ \cite{Mertes} 
and $V$ is the volume of the condensate.   
The rates of two-body atom losses depend on the density of the
atomic gases. We assume that $V$ is about
$(1\mu{\rm m})^3$.
The rates of two-body atom loss range from 1 Hz to 10 Hz, for $N={10}$ to 100.
To obtain the good performance, the coupling strength of the 
magnetic-field gradient $\Omega_{D}$ must be much larger than
the two-body atom loss rates.  
The coupling strength $\Omega_{L(R)}$ between the two states is 
about $\mu_B{B}_{L(R)}/\hbar$ \cite{Bohi},
where $\mu_B$ is the Bohr magneton.
Thus, the minimum value of magnetic field can be detected ranging 
from $10^{-10}$T to $10^{-9}$T, for $N=10$ to 100 and $V=(1\mu{\rm m})^3$.
The minimum value of the detectable magnetic-field gradient 
is about $10^{-9}{\sim}10^{-10}$T.

\section{Conclusion}
In summary, we have proposed a method to 
detect the magnetic-field gradient by using entangled
condensates.  We have described how to 
generate entangled states of two spatially separated 
condensates.  The magnetic-field gradient can be determined
by measuring the variances of relative phases and population 
differences between the two-component condensates in the two
wells.  The uncertainty of the parameter scales with $1/N$. 
We have also numerically studied the
effects of one-body and two-body atom losses
on the detection.  We show that the entangled atoms can 
outperform the uncorrelated atoms in detecting the magnetic
fields for a few atoms.  The effect of 
atom-atom interactions on this method has also
been discussed.

\begin{acknowledgments}
We thank Shih-I Chu.
This work was supported in part by the National Basic Research Program 
of China Grant 2011CBA00300, 2011CBA00301 and
the National Natural Science Foundation 
of China Grant 61073174, 61033001, 61061130540.
\end{acknowledgments}

\end{document}